\documentclass[aps,prb,twocolumn,showpacs]{revtex4-1}
\usepackage{graphicx}% Include figure files
\usepackage{natbib}
\usepackage[utf8]{inputenc}
\usepackage{epstopdf}
\setcitestyle{super}
\usepackage{color}
\usepackage[normalem]{ulem}
\usepackage{pdfpages}
\usepackage{hyperref}
\makeatletter
\AtBeginDocument{\let\LS@rot\@undefined}
\makeatother

\newcommand*{\citen}[1]{%
  \begingroup
    \romannumeral-`\x % remove space at the beginning of \setcitestyle
    \setcitestyle{numbers}%
    \cite{#1}%
  \endgroup   
}

\usepackage{hyphenat}
\begin{document}

\preprint{APS/123-QED}
\title{Multiple ferromagnetic transitions and structural distortion in the van-der-Waals ferromagnet VI$_3$ at ambient and finite pressures}
\author{Elena Gati$^{1,2}$}
\author{Yuji Inagaki$^{1,2,3}$}
\author{Tai Kong$^{4}$}
\author{Robert J. Cava$^{4}$}
\author{Yuji Furukawa$^{1,2}$}
\author{Paul C. Canfield$^{1,2}$}
\author{Sergey L. Bud'ko$^{1,2}$}
\address{$^{1}$Ames Laboratory, US Department of Energy, Iowa State University, Ames,
Iowa 50011, USA}
\address{$^{2}$Department of Physics and Astronomy, Iowa State University, Ames, Iowa 50011, USA}
\address{$^{3}$Department of Applied Quantum Physics, Faculty of Engineering, Kyushu University, Fukuoka 819-0395, Japan}
\address{$^{4}$Department of Chemistry, Princeton University, Princeton, New Jersey 08544, USA}

\date{\today}

\begin{abstract}
We present a combined study of zero-field $^{51}$V and $^{127}$I NMR at ambient pressure and specific heat and magnetization measurements under pressure up to 2.08\,GPa on bulk single crystals of the van-der-Waals ferromagnet VI$_3$. At ambient pressure, our results consistently demonstrate that VI$_3$ undergoes a structural transition at $T_s\,\approx\,$78\,K, followed by two subsequent ferromagnetic transitions at $T_{FM1}\,\approx\,$50\,K and $T_{FM2}\,\approx\,$36\,K upon cooling. At lowest temperature ($T\,<\,T_{FM2}$), two magnetically-ordered V sites exist, whereas only one magnetically-ordered V site is observed for $T_{FM1}\,<\,T\,<\,T_{FM2}$. Whereas $T_{FM1}$ is almost unaffected by external pressure, $T_{FM2}$ is highly responsive to pressure and merges with the $T_{FM1}$ line at $p\,\approx\,0.6\,$GPa. At even higher pressures ($p\,\approx\,$1.25\,GPa), the $T_{FM2}$ line merges with the structural transition at $T_s$ which becomes moderately suppressed with $p$ for $p\,<\,1.25$\,GPa. Taken together, our data point towards a complex magnetic structure and an interesting interplay of magnetic and structural degrees of freedom in VI$_3$.
\end{abstract}

\pacs{xxx}

\maketitle

\section{Introduction}

The search for two-dimensional (2D) materials with novel, highly-tunable electronic ground states is stimulated by their great potential for future device applications \cite{Novoselov05,Butler13,Das15}. In particular for spintronics, 2D insulating or semiconducting ferromagnets with elevated Curie temperatures, $T_c$, are desired \cite{Park16,Samarth17}. However, low-dimensional magnets are rare for fundamental reasons: thermal fluctuations destroy any long-range order in isotropic systems in one or two dimensions at any finite temperature, $T$, according to the Mermin-Wagner theorem \cite{Mermin66}. Thus, finite-$T$ magnetism in a low-dimensional system can only be achieved in systems with magnetic anisotropy. If identified, such anisotropic low-dimensional systems might also serve as a solid-state realization of well-established theoretical models \cite{Burch18,Enoki03,Dresselhaus92}. Examples include the Ising or the XY-model, which feature transitions, such as the Berezinskii–Kosterlitz–Thouless transition (BKT transition) \cite{Berezinskky71,Kosterlitz73}. 

Recently, magnetic van-der-Waals (vdW) materials were introduced as promising candidates for truly low-dimensional magnetism \cite{Park16,Burch18} since their vdW nature suggests the possibility of exfoliation down to the monolayer level and their low crystalline symmetry implies an intrinsic magnetic anisotropy. Prominent ferromagnetic members of this material class include CrI$_3$ ($T_c\,\approx\,61$K) \cite{Hansen59,Dillon65,McGuire15,Liu18}, CrBr$_3$ ($T_c\,\approx\,37$\,K) \cite{Tsubokawa60}, CrSiTe$_3$ ($T_c\,\approx\,33\,$K) \cite{Carteaux91} and Cr$_2$Ge$_2$Te$_6$ ($T_c\,\approx\,61\,$K) \cite{Carteaux95} (with the latter two being stable in air). Indeed, it was possible to retain ferromagnetism in monolayers of CrI$_3$ \cite{Huang17} and bi-layers of Cr$_2$Ge$_2$Te$_6$ \cite{Gong17}, which were obtained by exfoliation of bulk single crystals. 

\begin{figure}
		\begin{center}
		\includegraphics[width=0.5\columnwidth]{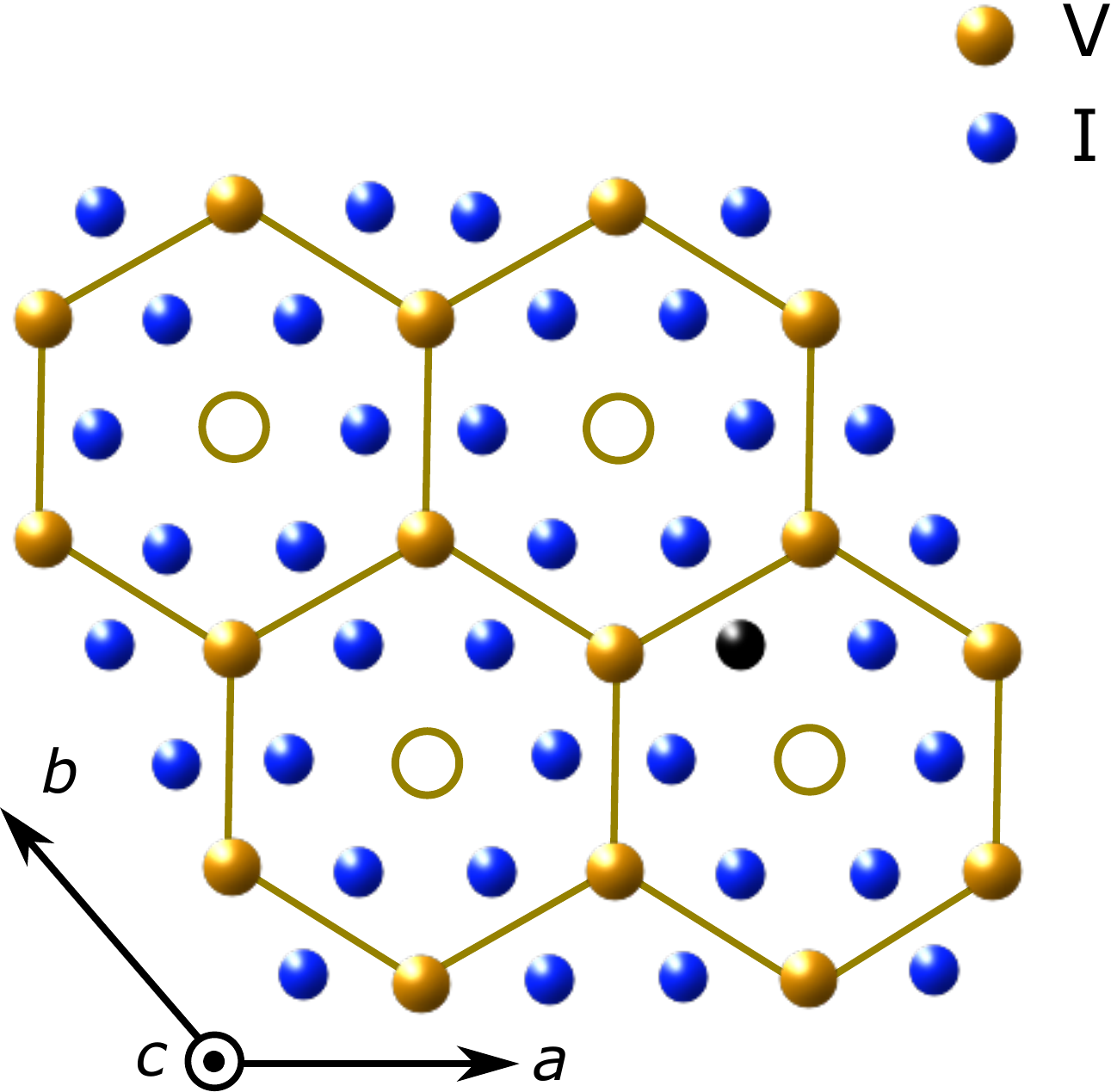} 
		\caption{Sketch of the in-plane crystal structure of VI$_3$ at room temperature: The V ions (orange), which are surrounded by I (blue), form a honeycomb lattice within the $ab$ plane. The open orange circles represent that x-ray measurements find a partial occupation of these V sites ($\approx\,4\%$) \cite{Kong19}, likely due to stacking faults in the structure. The magnetic easy axis was reported to be along the out-of-plane $c$ axis \cite{Kong19,Tian19,Son19}.}
		\label{fig:structure}
		\end{center}
		\end{figure}

Motivated by exploring the tunability of these vdW materials, there are ongoing efforts to identify new bulk ferromagnetic members of this material class with the potential for exfolation. As a result, ferromagnetism below $T_c\,\approx\,$50\,K was discovered in bulk single crystals of VI$_3$ in three very recent and almost simultaneous studies \cite{Kong19,Tian19,Son19}. Similar to other transition-metal trihalides \cite{McGuire15,Morosin64,Handy52,McCarley64}, such as CrI$_3$, VI$_3$ consists of stacked layers in which edge-sharing VI$_6$ octahedra form a honeycomb lattice (see Fig.\,\ref{fig:structure} for a sketch). Even though there is some disagreement between these three reports \cite{Kong19,Tian19,Son19} as to the detailed symmetry of the room-temperature structure, there is consensus that upon lowering $T$, VI$_3$ undergoes a structural transition at $T_s\,\approx\,79$\,K, i.e., at temperatures higher than the ferromagnetic ordering.

Experimentally, the properties of exfoliated VI$_3$ have not been reported to date. Meanwhile, a clear understanding of the bulk magnetic structure and its tunability can be considered as an important step towards predicting material's properties upon exfoliation. So far, the tunability of magnetism in VI$_3$ was tested on bulk single crystals by applying hydrostatic pressure \cite{Son19}. In these layered materials, hydrostatic pressure is expected to cause a strongly anisotropic modification of the intra- and inter-layer interactions which might therefore allow insight into their respective role for the magnetic ordering. For VI$_3$, it was reported \cite{Son19} that the Curie temperature is initially insensitive to pressure, but starts to increase abruptly with pressure above 0.6\,GPa. This behavior was interpreted as a crossover from two- to three-dimensional magnetism \cite{Son19}. However, the origin of this sudden change in the pressure dependence of the Curie temperature, as well as the potential connection to the structural transition have not yet been addressed.

In the present study, we examine the interplay of magnetic and structural degrees of freedom in VI$_3$ at ambient and finite pressures up to 2.08\,GPa via local (zero-field $^{51}$V and $^{127}$I NMR) and thermodynamic probes (specific heat and magnetization). From these combined measurements, we infer that VI$_3$ actually undergoes two magnetic transitions at ambient pressure (hereafter labeled as $T_{FM1}\,\approx\,51$\,K and $T_{FM2}\,\approx\,36$\,K, respectively). The resulting two magnetic phases (FM2 for $T\,<\,T_{FM2}$ and FM1 for $T_{FM2}\,<\,T\,<\,T_{FM1}$) at ambient pressure are characterized by a ferromagnetic component, and as such, we will label both transitions as ferromagnetic transitions throughout the entire manuscript (even though each state may well also have some finite ordering wave-vector, $q$, as well). In the FM2 state, two distinct, ordered V sites exist, whereas only one ordered V site is observed in the FM1 state. This suggests a complex magnetic ordering at low temperatures. Upon pressurization, we find two, well separated, triple points in the phase diagram, (i) $(p_{c1},T_{c1})\,=\,(0.6\,\textnormal{GPa}, 50.8\,\textnormal{K})$ at which the $T_{FM2}$ line merges with the $T_{FM1}$ line and (ii) $(p_{c2},T_{c2})\,=\,(1.25\,\textnormal{GPa}, 61.6\,\textnormal{K})$ at which the $T_{FM2}$ line merges with the $T_s$ line. Taken together, our results therefore point towards an interesting interplay of magnetic and structural degrees of freedom in VI$_3$. This conclusion emphasizes the need for careful scattering experiments in all salient temperature and pressure regions of the phase diagram for a clarification of the crystallographic structure of this material. This input is needed for the ultimate identification of the magnetic structures of VI$_3$.

\section{Experimental Details}
\label{sec:expdetails}

Bulk single crystals of VI$_3$, used in this study, were synthesized using chemical-vapor transport. The detailed procedure is described in Ref.\,\citen{Kong19}. Since these crystals are sensitive to humidity, all preparation work for the experiments was performed in a N$_2$ glovebox and crystals were only exposed to air for a very short time while transferring to the cryostat (in case of NMR measurements) or the pressure cell (for specific heat or magnetization measurements).

Specific heat measurements under pressure were performed on a single crystal of dimensions $\approx\,1.5\,\times\,1\,\times\,0.1\,$mm$^3$ using the technique of $ac$ calorimetry. In $ac$ calorimetric measurements, the sample is heated by an oscillating power with frequency $\omega$, and the resulting temperature oscillation, which is directly related to the specific heat of the sample, can be measured with high precision using a lock-in amplifier. By tuning the frequency of this $ac$ modulation, the sample can be decoupled, to a good approximation, from the pressure cell enviroment. The determination of the optimal frequency is therefore crucial for the present measurements. The optimal frequency depends on the thermal conductances of the specific arrangement, given by the sample's intrinsic thermal conductivity, its thickness, as well as the used glue, which is used for ensuring thermal contact between heater, sample and thermometer, as well as by the pressure medium. For the present experiments (on a sample with thickness of $\approx\,\,100\,\mu$m and glue thickness of $<\,30\,\mu$m), the optimal frequency was determined experimentally prior to measurements of the specific heat. It typically ranged from 200\,Hz at base temperature to 2\,Hz at 100\,K. The detailed measurement protocol and more details of the setup are described in Ref.\,\citen{Gati19}. A hybrid piston-pressure cell, made out of Grade 5 titanium alloy (Ti 6Al-4V; outer cell body) and Ni-Cr-Al alloy (inner cell body) \cite{Budko84}, was used to apply pressure up to 2.08\,GPa. A 4:6 mixture of light mineral oil:$n$-pentane was used as a pressure-transmitting medium. It solidifies at room temperature in the pressure range 3-4\,GPa, which is well above the maximum pressure of the setup and thereby offers good hydrostatic pressure conditions in the full pressure range investigated\cite{Budko84,Kim11,Torikachvili15}. Specific heat data were obtained in an increasing pressure cycle. Pressure values in the manuscript correspond to the pressures at low temperatures, which were determined from the pressure dependence of the critical temperature, $T_c(p)$, of elemental Pb from resistance measurements \cite{Bireckoven88,Eiling81}. Measurements in zero field at all pressures were performed in a cryogen-free cryostat from Janis (SHI-950). At highest pressure, additional measurements in zero and applied magnetic field were performed in a cryogen-free cryostat from ICEOxford (Lemon $^{\textnormal{DRY}}$ICE$^{\textnormal{NS-TL50}}$) with a maximum field of $\mu_0 H\,=\,9\,$T.

For magnetization measurements under pressure, an aggregate of single crystals with random orientation was placed in the pressure cell. Due to the large mass of crystals, the signal of the samples is large compared to the weakly temperature-dependent background signal from the cell. As a result, the cell contribution can be considered negligible to a good approximation. Given that the present study focusses on the determination of transition temperatures, we did not obtain a high-accuracy measure of the weight and therefore we omit a normalization and conversion of magnetization into susceptibility in the pressure data sets. The pressure cell used in this study is a commercially-available HDM pressure cell \cite{QDcell} with maximum-available pressure of $\approx\,1\,$GPa. Good hydrostatic pressure conditions were provided by using Daphne 7373 as a pressure-transmitting medium which solidifies at room temperature at 2.2\,GPa \cite{Yokogawa07}. Magnetization data were taken upon increasing and decreasing pressure. No qualitative difference between measurements taken upon increasing and decreasing pressure was found. This indicates that the orientation of the individual crystals in the aggregate was not changed significantly throughout the pressure cycle.  Superconducting Pb was again used as a pressure gauge at low temperatures \cite{Bireckoven88,Eiling81}. Measurements were performed in a Quantum Design Magnetic Property Measurement System (MPMS-3) SQUID magnetometer. Temperature-dependent data sets were collected in low fields ($\mu_0 H\,=\,2\,$mT) and high fields ($\mu_0 H\,=\,0.1\,$T) after zero-field cooling. As the present study primarily focuses on the determination of transition temperatures (due to the complications involved which arise from the random orientation of the crystals), we restricted the study to measurements after zero-field cooling. For a detailed comparison of zero-field and field-cooled measurements, we refer to the data of Refs.\,\cite{Kong19,Tian19,Son19} at ambient pressure on one single crystal, which was well aligned in magnetic field.

The error of the pressure determination from the shift of the $T_c$ of Pb typically amounts to $\pm 0.02$\,GPa. In addition, pressure is subject to temperature changes; however, below $\approx\,100\,$K, for which the pressure medium is solidified, pressure is almost constant with $\Delta p\,<\,0.03$\,GPa between low temperature and 100\,K (i.e., the temperature range of interest in the present work).

NMR  measurements of $^{51}$V ($I$ = $\frac{7}{2}$, $\frac{\gamma_{\rm N}}{2\pi}$ = 11.193 MHz/T, $Q=$ -0.052 barns) and  $^{127}$I  ($I$ = $\frac{5}{2}$, $\frac{\gamma_{\rm N}}{2\pi}$ = 8.557 MHz/T, $Q=$ 0.09298 barns) nuclei in VI$_3$ were conducted using a laboratory-built, phase-coherent, spin-echo pulse spectrometer. For this purpose, the plate-like single crystals, with typical in-plane dimensions of less than $1\,\times\,1\,$mm$^2$, were loosely packed in a several mm wide NMR sample capsule. By applying a magnetic field of 7\,T, which is well above the saturation field for VI$_3$ \cite{Kong19,Tian19,Son19}, while cooling through the ferromagnetic transition, the crystals inside the capsule are encouraged to align so that the easy axis points in the direction of the external magnetic field. As we will show below, our NMR results indeed speak in favor of a preferential orientation of the single crystals in the NMR capsule. The $^{51}$V and $^{127}$I zero-field NMR spectra were obtained by sweeping the frequency at zero magnetic field in the ferromagnetic state. Attempts to measure $^{51}$V NMR and $^{127}$I NMR in the paramagnetic state were performed, but no NMR signals were observed. The $^{51}$V transverse  relaxation time  ($T_2$) at each temperature ($T$) was determined by fitting the nuclear magnetization $M$ versus time $2\tau$  using the exponential function $M(2\tau) = M(0)e^ {-2\tau/T_{2}}$ where $M(2\tau)$ is the nuclear magnetization at $2\tau$ after the application of $\pi$/2-$\pi$ radio frequency pulses separated by $\tau$.

\section{Results}

	\subsection{Magnetism at ambient pressure}
	
	\begin{figure}
		\begin{center}
		\includegraphics[width=\columnwidth]{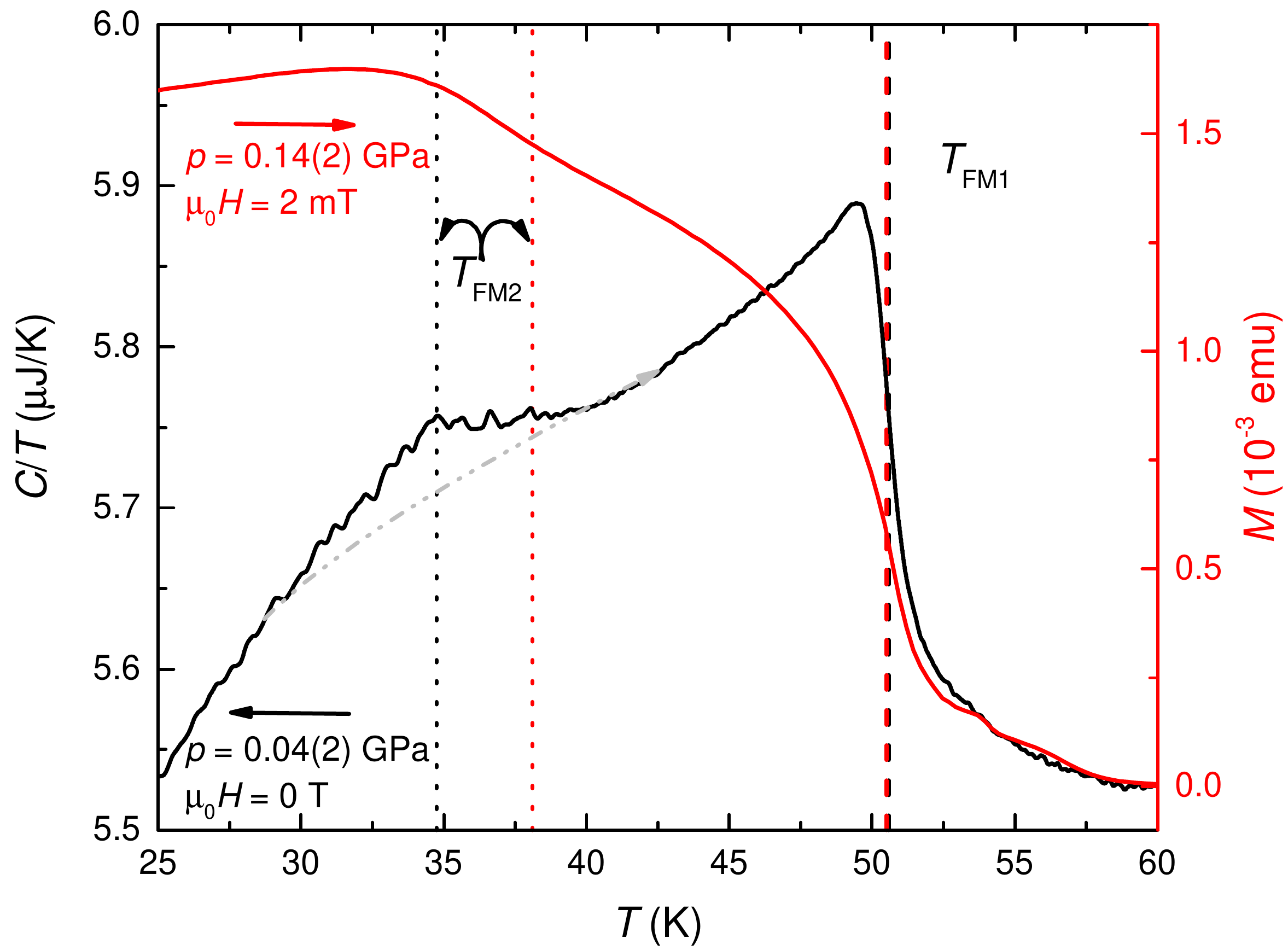} 
		\caption{Comparison of specific heat, $C/T$ (black line, left axis), and magnetization, $M$ (red line, right axis), as a function of temperature at lowest pressures measured (0.04\,GPa and 0.14\,GPa, respectively). Dashed (dotted) lines correspond to the position of $T_{FM1}$ ($T_{FM2}$), determined from $C/T$ and $M$. The discrepancy in $T_{FM2}$ values is likely related to the small pressure difference between the $C/T$ and $M$ data sets which affects $T_{FM2}$ stronger than $T_{FM2}$ due to their different responses to pressure (see main text). To better visualize the broad peak shape of the FM2 feature in $C/T$ (the maximum of which is marked by the black dotted line), a potential background contribution is given by the grey dashed-dotted line.}
		\label{fig:ambient-thermodynamics}
		\end{center}
		\end{figure}
		
		First, we focus on the magnetic properties of VI$_3$ at (or very close to)  ambient pressure. In Fig.\,\ref{fig:ambient-thermodynamics}, we show thermodynamic data of the magnetization, $M$, and specific heat, $C/T$, below $T\,\le\,60\,$K (Data for higher $T$, showing $T_s$, are presented in Fig.\,\ref{fig:CMlowp}, below). These two data sets were taken in the pressure-cell environment (so as to protect the sample from humidity or degradation) with small externally applied force (hand-tight nuts), resulting in a small, but finite pressure at low temperatures. This pressure can be quantified by using a Pb manometer, and amounts to 0.14\,GPa for the $M$ data and 0.04\,GPa for the $C/T$ data. 
		
		Upon lowering the temperature, $C/T$ shows a clear $\lambda$-type feature at $T_{FM1}\,\approx\,51\,$K, which is accompanied by a steep increase of $M$. The given $T_{FM1}$ value corresponds to the temperature, at which the temperature-derivative of the $C/T$ and $M$ data sets shows a pronounced minimum (see Section B for a more detailed discussion of criteria used to infer $T_{FM1}$ and $T_{FM2}$). From these observations, we assign $T_{FM1}$ as the transition from a paramagnetic to a magnetically-ordered state, consistent with all previous reports \cite{Kong19,Tian19,Son19}. The $\lambda$-type shape of the specific heat feature speaks in favor of a mean-field type second-order phase transition. At lower temperatures, $C/T$ shows a second, more subtle, feature at $\,\approx\,35$\,K (determined from the mid-point of a step-like feature in d($C/T$)/d$T$). Here, a small, more symmetric peak is observed in $C/T$ on top of the dominant background specific heat. Very close to this temperature, at $T\,\approx\,$38\,K (corresponding to the minimum temperature in d$M$/d$T$), $M$ shows a subsequent strong increase to lower $T$. We thus assign these features to a second ferromagnetic transition occurring in VI$_3$ at $T_{FM2}\,\approx\,(36\,\pm\,2)$\,K at ambient pressure. The small discrepancy in transition temperatures likely originates from a combination of the little pressure difference between the $M$ and $C/T$ measurements and the high pressure sensitivity of this transition (see Section B). 
		
		This second ferromagnetic transition at lower $T$ at ambient pressure was not identified in a previous work \cite{Kong19,Tian19,Son19}, as the specific heat does not show a pronounced feature at this transition. This specific heat feature, however, becomes significantly enhanced and sharper at finite pressures, as we will discuss below. This finding will strengthen our conclusion further and will then also allow us to comment on the order of the phase transition at $T_{FM2}$.

   		\begin{figure}
		\begin{center}
		\includegraphics[width=\columnwidth]{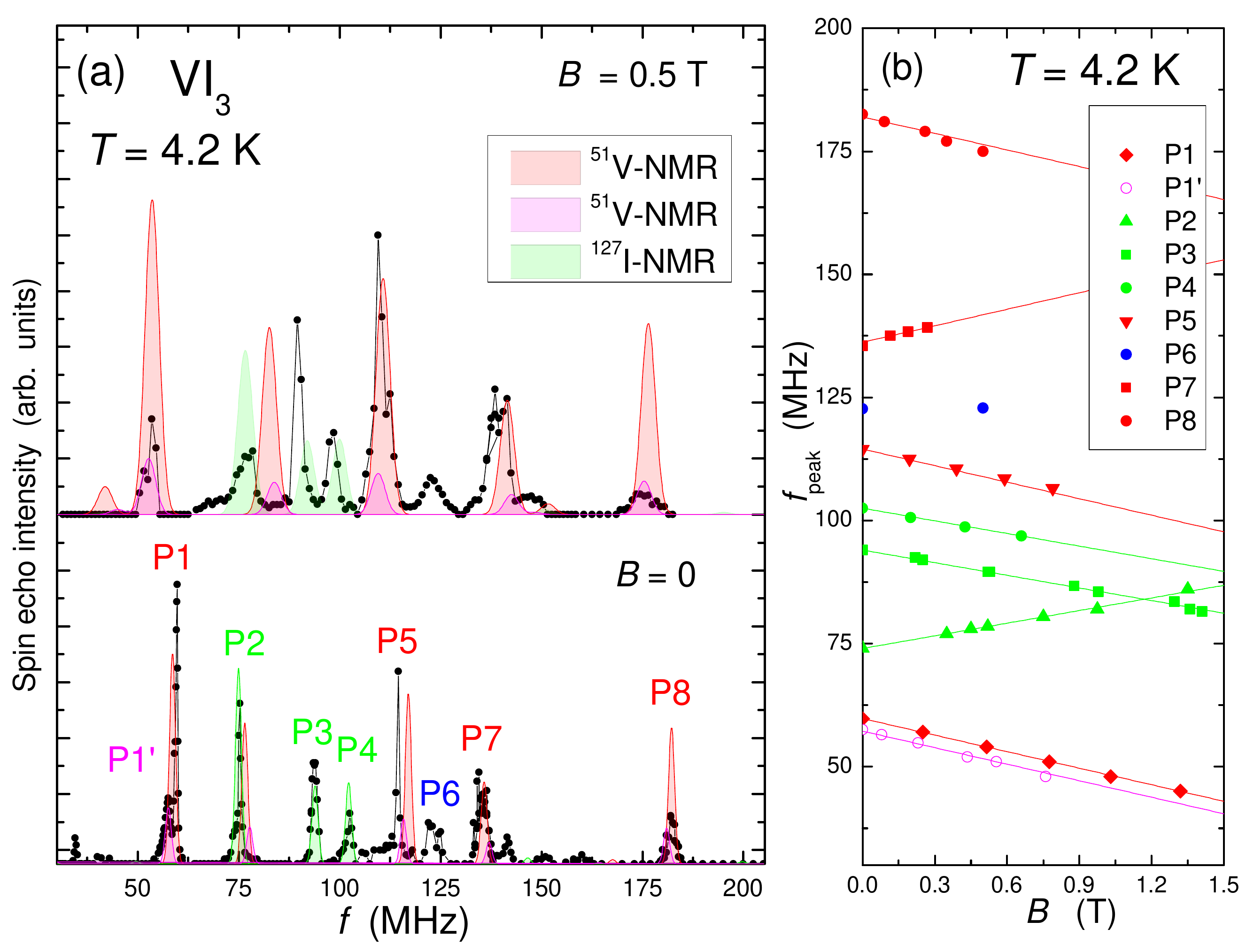} 
		\caption{(a) NMR spectrum, measured in zero magnetic field (bottom) and magnetic field of 0.5\,T (top), at $T$ = 4.2 K in the ferromagnetic state for VI$_3$ (black circles). In total, nine major peaks are observed, which are labeled with P1, P1$^\prime$ and P2 to P8. Colored lines represent simulated NMR spectra for two $^{51}$ V sites (red and pink) as well as for a $^{127}$I site. The set of parameters used for the simulations are denoted in the main text. The same parameters were used for the simulations in the top and bottom panel; (b)  Magnetic field dependencies of the resonance frequency ($f_{\rm peak}$) for each peak of the major peaks P1 to P9, measured at $T$ = 4.2 K. The absolute value of slopes for the red and pink solid line is 11.193\,MHz/T corresponding to $\frac{\gamma_{\rm N}}{2\pi}$  for the $^{51}$V nucleus, while that for the green lines is 8.557 MHz/T corresponding to the $\frac{\gamma_{\rm N}}{2\pi}$ value of the $^{127}$I nucleus.}
		\label{fig:NMRspectrum1}
		\end{center}
		\end{figure}

		\begin{figure*}
		\begin{center}
		\includegraphics[width=\textwidth]{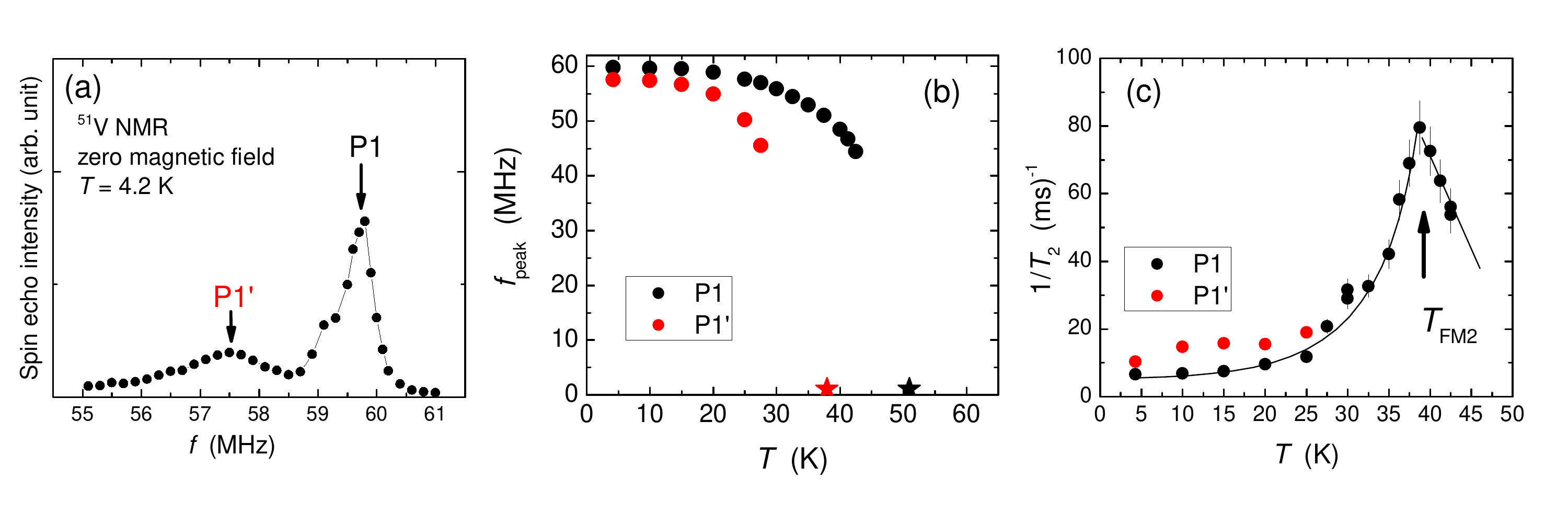} 
		\caption{(a) NMR spectrum in VI$_3$ measured at $B$ = 0\,T and $T$ = 4.2 K at ambient pressure in the ferromagnetic state. The two peaks (labelled by P1 and P1$\prime$ correspond to central transition lines of $^{51}$V NMR. The ratio of the integrated intensities of P1 and P1$^\prime$ is estimated to be 1.7 : 1 after correction by their respective longitudinal and transverse relaxation times ($T_1$ and $T_2$, respectively); (b) Temperature dependence of $f_{\rm peak}$ for each peak at zero magnetic field. Black (red) star denotes the transition temperature  $T_{FM2}$ ($T_{FM1}$), inferred from specific heat and magnetization data in Fig.\,\ref{fig:ambient-thermodynamics}; (c) Temperature dependence of 1/$T_2$ for both P1 and P1$^\prime$ at zero magnetic field. 1/$T_2$ shows a peak at $T\,\sim\,$38 K, which is very close to the second ferromagnetic phase transition temperature $T_{\rm FM2}$, determined from magnetization and specific heat measurements. The solid curve is a guide to the eye.}
		\label{fig:NMRspectrum}
		\end{center}
		\end{figure*}

 To confirm the existence of the two ferromagnetic phase transitions  from a microscopic point of view,  we carried out NMR measurements at ambient pressure. The bottom of Fig.\,\ref{fig:NMRspectrum1}(a) shows the NMR spectrum recorded in zero magnetic field at $T\,$=\,4.2 K in the FM2 state. Several peaks occur in the frequency range 35\,MHz$\,\le\,f\,\le$\,200\,MHz as a result of a superposition of $^{51}$V and $^{127}$I NMR signals. To assign the peaks to $^{51}$V and $^{127}$I signals, we measured the magnetic field ($B$) dependence of the frequencies of the peak positions ($f_{\rm peak}$) for the nine major peaks, labeled with P1, P1$^\prime$ and P2 to P8 in Fig. \,\ref{fig:NMRspectrum1}(a) (see discussion below for a detailed explanation of the peak notation). Figure\,\ref{fig:NMRspectrum1}\,(b) shows that, with increasing $B$, the majority of peaks shift to lower frequency, while the peaks P2 and P7 clearly shift to higher frequency and P6 does not show any significant shift. From the slope of the variation of $f_{\rm peak}$ with $B$, which is specific to the respective NMR nucleus, we infer that P2, P3 and P4 originate from $^{127}$I NMR signals, whereas the others (with the exception of P6) can be assigned to $^{51}$V NMR signals. 
 
The fact that all peaks (except for P6) show net shifts in small applied fields, which are lower than the anisotropy field of $\sim$ 1 T at $T\,=\,$4.2 K, and that the shifts can be consistently explained by the gyromagnetic ratios of $^{51}$V and $^{127}$I, indicates that the major part of the single crystals in the aggregate are oriented with the easy axis parallel to the external magnetic field. As outlined in detail in Section \ref{sec:expdetails}, this is a result of the loose packing of the small single crystals together with the application of a large field of 7\,T prior to the measurements. In case of a random orientation of single crystals, one would not expect any net shift of the peak positions, but rather a significant broadening of the peaks.

    To understand the measured spectrum more quantitatively, we simulated the spectrum based on the following nuclear spin Hamiltonian ${\cal H} = {\cal -}\gamma\hbar{\bf I}\cdot{\bf B_{\rm eff}}+ \frac{h \nu_{\rm Q}}{6} [3I_{z}^{2}-I^2 + \frac{1}{2}\eta(I_+^2 +I_-^2)]$. Here, $B_{\rm eff}$ is the effective field at the nuclear site, given by the sum of the internal magnetic field $B_{\rm int}$ and the external magnetic field $B$, $h$ is Planck's constant, and $\nu_{\rm Q}$ is the nuclear quadrupole frequency. The latter is defined by $\nu_{\rm Q} = 3e^2QV_{ZZ}/2I(2I-1)h$,  where $Q$ is the quadrupole moment of nucleus, $V_{ZZ}$ is the electric field gradient (EFG) at the nuclear site, and $\eta$ is the asymmetry parameter of the EFG. As shown in the bottom of Fig. \,\ref{fig:NMRspectrum1}(a),  the observed spectrum under zero magnetic field is relatively well reproduced by the simulation. The red (pink) lines represent the calculated $^{51}$V NMR spectrum using a set of parameters of $B_{\rm int}$ = -4.90 T ($B_{\rm int}\,=\,$-4.70 T), $\nu_{\rm Q}$ = 63.8 MHz ($\nu_{\rm Q}$ = 63.8 MHz), $\eta$ = 0.132 ($\eta$ = 0.132)  and $\theta = 0^\circ$ ($\theta = 0^\circ$)  and  the green lines represent the $^{127}$I NMR spectrum using $B_{\rm int}$ = -1.2 T, $\nu_{\rm Q}$ = 49.9 MHz, $\eta$ = 0.91 and  $\theta = 0^\circ$. Here $\theta$ represents the angle between $B_{\rm int}$  and the principle axis of the EFG tensor at each nuclear site. We used the same set of parameters to calculate the spectrum under $B$ = 0.5 T for $^{51}$V  and $^{127}$I nuclei. The calculated spectrum reproduces the measured one to a good approximation, as shown at the top of Fig. \,\ref{fig:NMRspectrum1}(a), although a few of the peaks cannot be perfectly explained by the simulation. The origin for the additional peaks, like P6, is unclear at present. One possible source for these additional peaks might be small impurity phases in the crystal, potentially as a result of their sensitivity to humidity. Another possibility might be related to stacking faults in these layered materials, which cause a minor part of V sites to have a different local magnetic environment. In fact, recent x-ray data \cite{Kong19} found a partial occupancy of $4\%$ of nominally vacant interstitial lattice sites, which were attributed to the presence of stacking faults.
    
   The finite $B_{\rm int}$ values for the $^{51}$V and $^{127}$I NMR peaks demonstrate the existence of static hyperfine field at both nuclear sites produced by the ordered V moments, confirming the ordered state with ferromagnetic components. The negative sign of the internal fields at the V site suggests a dominant contribution of 3$d$ electron core polarization to the total hyperfine field, whereas the internal field at the I site originates from transferred hyperfine field produced by the V 3$d$ ordered moments. 
It should be noted though that the observed value of $B_{\rm int}$ at the V site is much smaller than those observed in several magnetic vanadium oxide-based compounds (for example, $B_{\rm int}$ = -24.91 T in YVO$_3$ \cite{Kikuchi94}, $B_{\rm int}$ = -23.72 T in LaVO$_3$ \cite{Kikuchi94} and $B_{\rm int}$ = -21.1 T in CaV$_2$O$_4$ \cite{Zong08}). At present, the reason for the small internal field at the V sites in VI$_3$ is not clear.  One possible explanation is that positive contributions of orbital and/or dipolar hyperfine couplings partially cancel the negative core polarization hyperfine field, as discussed previously\cite{Tsuda68,Fukai96,Roy13,Wiecki15}.

In the following, we focus on the peaks, which correspond to the to the central transition line ($I_z$ = $-$1/2$\leftrightarrow$1/2) of $^{51}$V NMR for the V site. From the analysis and the simulations above, we identify the peak around 59.5 MHz (P1)  in zero field at $T$ = 4.2 K to this central line (see Fig.\,\ref{fig:NMRspectrum}\,(a)). In addition, we observed a second, slightly broader and somewhat smaller peak around 57.5 MHz (labeled with P1$^\prime$), which can also be assigned to a central transition line of $^{51}$V NMR spectrum. The ratio of the integrated intensities of P1 and P1$^\prime$ is estimated to be 1.7:1. This estimate was obtained by correcting the intensities by their respective longitudinal and transverse relaxation times $T_1$ and $T_2$, respectively. The observation of two central $^{51}$NMR lines indicates that two V sites with ordered moments exist at low temperatures in VI$_3$.  It is noted that, although the central transition line around 57.5 MHz is well resolved, most of other transition lines of this second V site at higher frequencies cannot be clearly resolved, as they likely overlap with other signals and their signals are weaker in intensity.

With increasing temperature, both central line peaks shift to lower frequency (see Fig. \,\ref{fig:NMRspectrum}\,(b)). The corresponding decrease of $f_{\rm peak}$ is more rapid for P1$^\prime$ than for P1. Whereas the peak P1 can be resolved up to $\sim$\,43 K, the peak P1' becomes indiscernible already at lower temperatures (around 25 K). In general, the temperature dependence of $f_{\rm peak}$ reflects that of $B_{\rm int}$, which, in turn, is proportional to the spontaneous magnetization in the ferromagnetic state. Therefore, the $B_{int}$ values for P1 and P1$^\prime$ do not obey the same temperature dependencies. In particular, our data suggest distinctly different onset temperatures for $B_{int}$ for P1 and P1$^\prime$. As suggested by the black and red stars in Fig.\,\ref{fig:NMRspectrum}\,(b), which mark the transition temperatures $T_{FM1}$ and $T_{FM2}$, determined from specific heat and magnetization in the present study, it seems natural to infer that peak P1 sets in below $T_{FM1}$, whereas P1$^\prime$ occurs below $T_{FM2}$. Taking all the experimental observations together, we conclude that two different, ordered V sites exist in FM2, whereas there is only one ordered V site in FM1. Remarkably, $f_{peak}$ of P1 appears to be unchanged across $T_{FM2}$, indicating that there is no change of hyperfine field on this site while ordering on the second site occurs.

It is important to point out that measurements of $T_2$ (see Fig. \,\ref{fig:NMRspectrum}(c)) provide evidence that these two transitions reflect indeed the intrinsic properties of VI$_3$ and do not result from two different ferromagnetic phases with different $T_{FM}$. Up to 25\,K (the maximum temperature at which 1/$T_2$ for P1$^\prime$ was resolvable), 1/$T_2$ of P1 and P1$^\prime$ show a very similar temperature dependence. Upon further increasing the temperature, 1/$T_2$ of P1 shows a pronounced peak at 38\,K, i.e., at the second ferromagnetic transition temperature $T_{\rm FM2}$ which is associated with P1$^\prime$. If this second transition would originate from a spatially segregated, second magnetic phase, 1/$T_2$ of P1 would not be affected by the phase transition at $T_{\rm FM2}$. Therefore, these microscopic studies of the magnetism at ambient pressure on VI$_3$, together with our thermodynamic data, clearly suggest the existence of the two ferromagnetic phase transitions at $T_{\rm FM1}$ $\sim$ (51$\pm$1) K and $\sim$ (36$\pm$2) K  in VI$_3$ at ambient pressure.

	\subsection{Effect of pressure on magnetic and structural phase transitions}
	
\begin{figure*}
		\begin{center}
		\includegraphics[width=\textwidth]{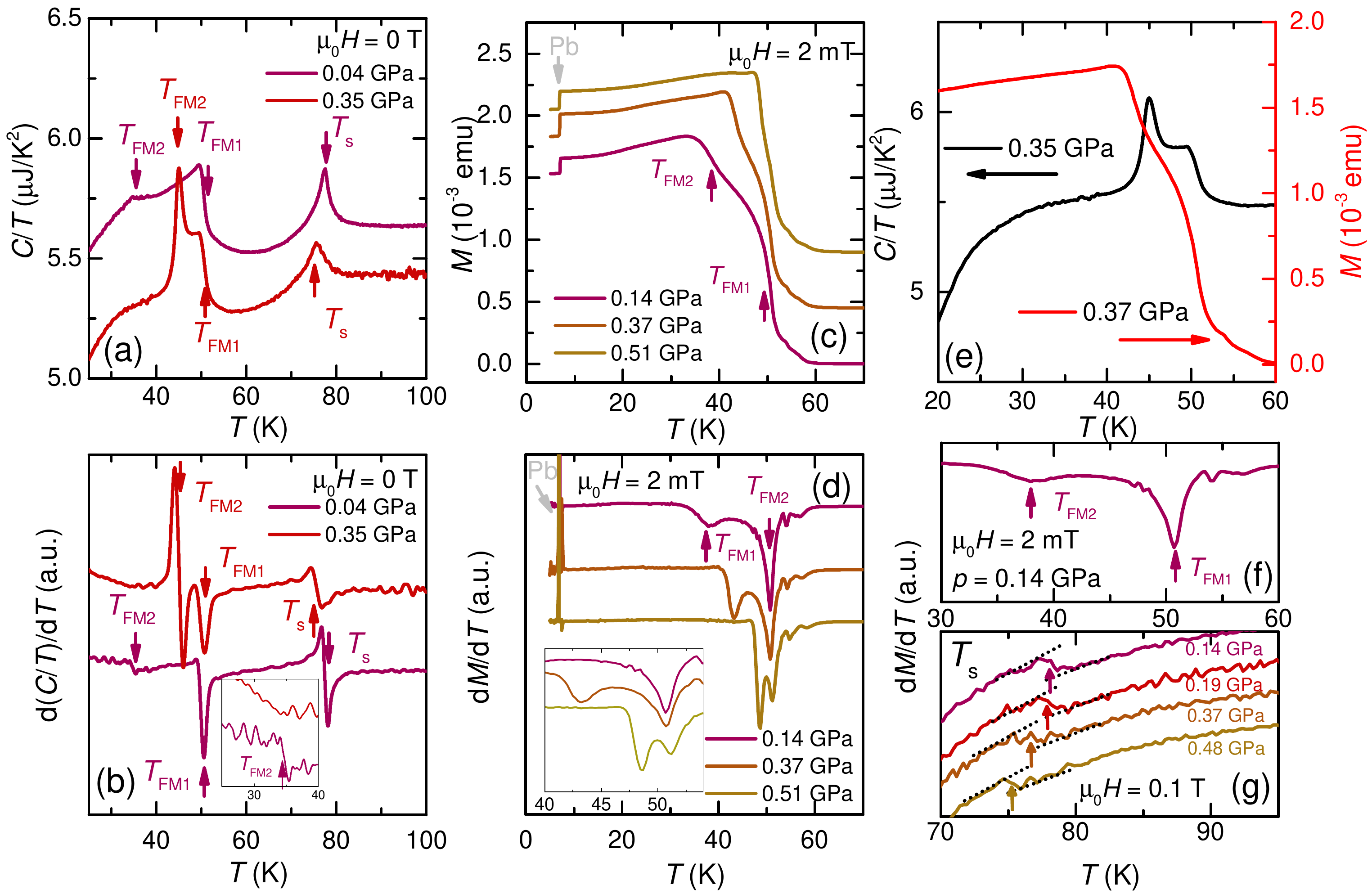} 
		\caption{Temperature-dependent specific heat, $C/T$, and magnetization data, $M$, of VI$_3$ at different pressures for $p\,\le\,0.51\,$GPa (Error for given pressure values is $\Delta p \,\approx\,\pm\,0.02$\,GPa.): (a) Specific heat data, taken in zero magnetic field. These data sets depict three distinct anomalies (marked by the arrows), which can be assigned to the structural transition at $T_s$ and the two ferromagnetic transitions at $T_{FM1}$ and $T_{FM2}$; (b) Temperature-derivative of the specific heat data, d$(C/T)$/d$T$, in (a). Arrows are used to visualize criteria to determine $T_{FM1}$, $T_{FM2}$ and $T_s$. The inset shows d$(C/T)$/d$T$ on enlarged scales around $T_{FM2}$ at $p\,=\,$0.04\,GPa; (c) Magnetization data, taken in a small field ($\mu_0$H\,=\,2\,mT). The position of the two ferromagnetic anomalies at $T_{FM1}$ and $T_{FM2}$ is marked by arrows. The step-like change of $M$ at $T\,\approx\,$7\,K can be assigned to the superconducting transition of elemental Pb, whch is used as a pressure manometer; (d) Temperature-derivative of magnetization data, d$M$/d$T$ in (a). The inset shows d$M$/d$T$ on enlarged scales; (e) Comparison of specific heat data, $C/T$, taken at 0.35\,GPa, and magnetization data, $M$, taken at 0.37\,GPa; (f) Temperature-derivative of magnetization, d$M$/d$T$, at 0.14\,GPa on enlarged scales. Arrows indicate criteria used to infer $T_{FM1}$ and $T_{FM2}$ from magnetization data; (g) Temperature-derivative of magnetization data, taken in a higher field of $\mu_0 H$\,=\,0.1\,T, around the structural transition at $T_s$. Arrow and dashed lines indicate the criterion to infer $T_s$. Data in (a)-(d) and (g) have been offset for clarity.}
		\label{fig:CMlowp}
		\end{center}
		\end{figure*}

\begin{figure*}
		\begin{center}
		\includegraphics[width=\textwidth]{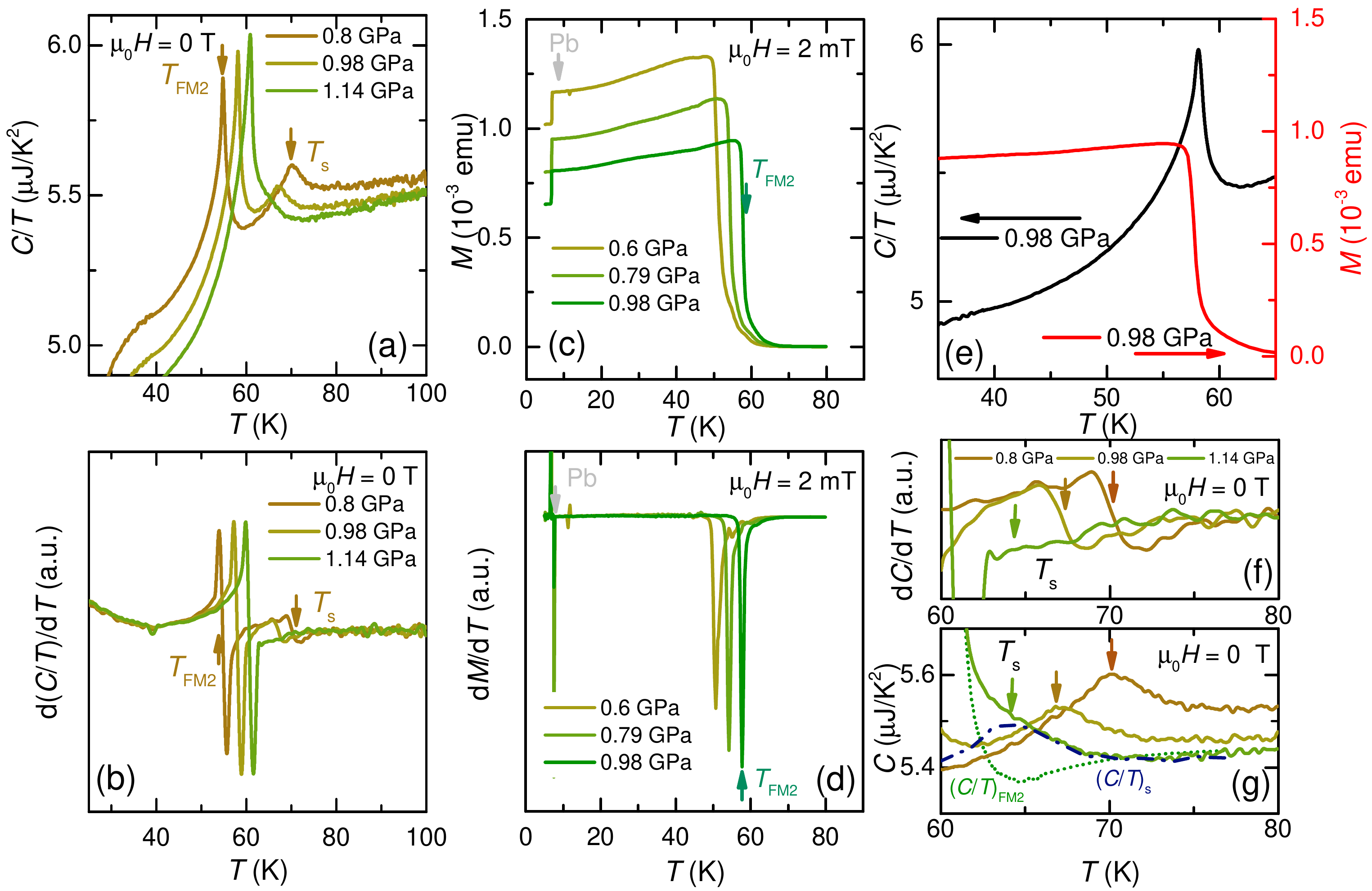} 
		\caption{Temperature-dependent specific heat, $C/T$, and magnetization data, $M$, of VI$_3$ at different pressures for 0.6\,GPa$\,\le\,p\,<\,1.35\,$GPa (Error for given pressure values is $\Delta p \,\approx\,\pm\,0.02$\,GPa.): (a) Specific heat data, taken in zero magnetic field. These data sets depict two distinct anomalies (marked by the arrows for $p\,=\,0.8$\,GPa), which can be assigned to the structural transition at $T_s$ and one ferromagnetic transitions at $T_{FM2}$; (b) Temperature-derivative of the specific heat data, d$(C/T)$/d$T$, in (a). Arrows are used to visualize criteria to determine $T_{FM2}$ and $T_s$ for $p\,=\,0.8$\,GPa; (c) Magnetization data, taken in a small field ($\mu_0$H\,=\,2\,mT). The position of the ferromagnetic anomaly at $T_{FM2}$ is marked by the arrow for $p\,=\,0.98\,$GPa. The step-like change of $M$ at $T\,\approx\,$7\,K can be assigned to the superconducting transition of elemental Pb, whch is used as a pressure manometer; (d) Temperature-derivative of magnetization data, d$M$/d$T$ in (a); (e) Comparison of specific heat data, $C/T$, taken at 0.98\,GPa, and magnetization data, $M$, taken at 0.98\,GPa; (f) Temperature-derivative of specific heat, d$(C/T)$/d$T$, at 0.8\,GPa, 0.98\,GPa and 1,14\,GPa on enlarged scales around $T_s$ (indicated by the arrows); (g) Temperature-dependent specific heat, $(C/T)$, at 0.8\,GPa, 0.98\,GPa and 1.14\,GPa on enlarged scales around $T_s$ (indicated by the arrows). The green dotted line shows an estimate of the specific heat peak, associated with $T_{FM2}$, $(C/T)_{FM2}$ at 1.14\,GPa. The dark blue dashed-dotted line, obtained by subtracting $(C/T)_{FM2}$ from the measured $C/T$, represents an estimate for the specific heat associated with the structural transition at $T_s$, $(C/T)_{s}$ (for details, see text).}
		\label{fig:CMmedp}
		\end{center}
		\end{figure*}

\begin{figure*}
		\begin{center}
		\includegraphics[width=\textwidth]{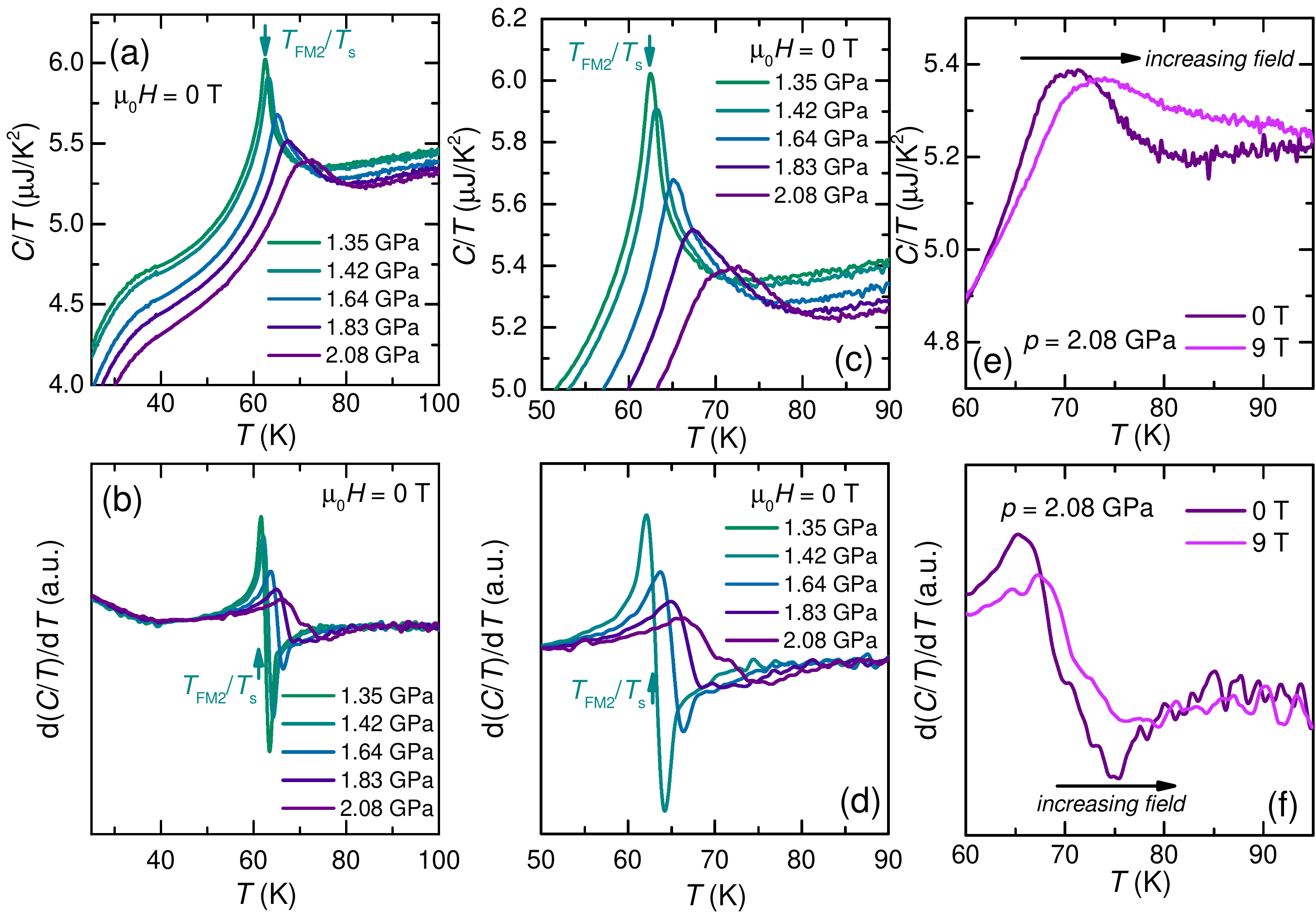} 
		\caption{Temperature-dependent specific heat, $C/T$, of VI$_3$ at different pressures for $p\,\ge\,1.35\,$GPa (Error for given pressure values is $\Delta p \,\approx\,\pm\,0.02$\,GPa.): (a) Specific heat data, taken in zero magnetic field. These data sets depict one anomaly (marked by the arrows for $p\,=\,1.35$\,GPa), which can be assigned to a simultaneous magnetic and structural transition at $T_{FM2}$/$T_s$; (b) Temperature-derivative of the specific heat data, d$(C/T)$/d$T$, in (a). Arrows are used to visualize criteria to determine $T_{FM2}$/$T_s$ for $p\,=\,1.3$\,GPa; (c,d) $C/T$ and d$(C/T)$/d$T$, shown in (a) and (b), on enlarged scales; (e) Temperature-dependent specific heat, $C/T$, at $p\,=\,2.08\,$GPa in zero field and applied field ($\mu_0$H\,=\,9\,T); (f) Temperature-derivative of specific heat, d$(C/T)$/d$T$, at 0.8\,GPa, 0.98\,GPa and 1.14\,GPa on enlarged scales around $T_s$ (indicated by the arrows); (g) Temperature-derivative of the data shown in (e).}
		\label{fig:CMhighp}
		\end{center}
		\end{figure*}

	In Figs.\,\ref{fig:CMlowp}-\ref{fig:CMhighp}, we show our results of specific heat, $C/T$, and magnetization, $M$, at finite pressures $p$. Importantly, the combination of these two thermodynamic probes allows us to trace not only the ferromagnetic transition lines $T_{FM1}(p)$ and $T_{FM2}(p)$ in the $T$-$p$ phase diagram, but also the structural transition line $T_s(p)$ as well.
	
	Before discussing the pressure-dependent data sets, we define the criteria which we used to determine $T_{FM1}$, $T_{FM2}$ and $T_s$ from the $C/T$ vs. $T$ and $M$ vs. $T$ data sets. To this end, we return to the data sets at lowest pressure (pressures close to ambient pressure), which are shown in Figs.\,\ref{fig:CMlowp}\,(a) and (c), respectively on a larger $T$ scale up to $T\,=\,$100\,K. In terms of the specific heat, the $\lambda$-type phase transition at $T_{FM1}$ results in a peak in the temperature derivative, d$(C/T)$/d$T$, shown in Fig.\,\ref{fig:CMlowp}\,(b). We assign the temperature of the minima (see arrow) of d$(C/T)$/d$T$ to $T_{FM1}$. The symmetric peak in $C/T$ at $T_{FM2}$ yields a step-like feature in d$(C/T)$/d$T$, with over- and undershoots at the low- and high-$T$ end of the phase transition (see inset of Fig.\,\ref{fig:CMlowp}\,(b) for d$(C/T)$/d$T$ on expanded scales around $T_{FM2}$ at 0.04\,GPa). Thus, we use the midpoint of this step-like feature to determine $T_{FM2}$ (see arrows in Fig.\,\ref{fig:CMlowp}\,(b)). Last, the structural transition, known from various ambient-$p$ studies \cite{Kong19,Tian19,Son19}, manifests itself as a slightly-broadened, symmetric peak in $C/T$ at higher temperatures ($T_s\,\approx\,78$\,K at $p\,=\,$0.04\,GPa). Here again, we use the midpoint of the step-like feature in d$(C/T)$/d$T$ to infer $T_s$ (see arrow in Fig.\,\ref{fig:CMlowp}\,(b)). In the magnetization data sets (see Fig.\,\ref{fig:CMlowp}\,(c)), the two ferromagnetic transitions each give rise to a steep increase of $M(T)$ upon lowering $T$. As a consequence, two minima are present in the derivative, d$M$/d$T$, as a function of $T$ (see Fig.\,\ref{fig:CMlowp}\,(d)). The position of these minima therefore define $T_{FM1}$ and $T_{FM2}$, as shown in Fig.\,\ref{fig:CMlowp}\,(f) for $p\,=\,0.14$\,GPa on expanded $T$ scales. In addition, a subtle step-like feature at high temperatures in d$M$/d$T$ (taken in $\mu_0 H\,=\,0.1\,$T, see Fig.\,\ref{fig:CMlowp}\,(g)), which can be associated with the structural transition at $T_s$, is observed. We assign the midpoint of the step-like feature to $T_s$. The fact that the structural transition gives rise to change of the magnetization in the paramagnetic state at high temperatures is an indication for modified spin interactions as a result of the structural distortion. Thus, this can be considered as a strong hint towards the significance of spin-lattice coupling in this compound \cite{McGuire15,Kong19,Tian19,Son19}, which therefore should be included in discussion of its magnetic properties.
	
\begin{figure}
		\begin{center}
		\includegraphics[width=0.95\columnwidth]{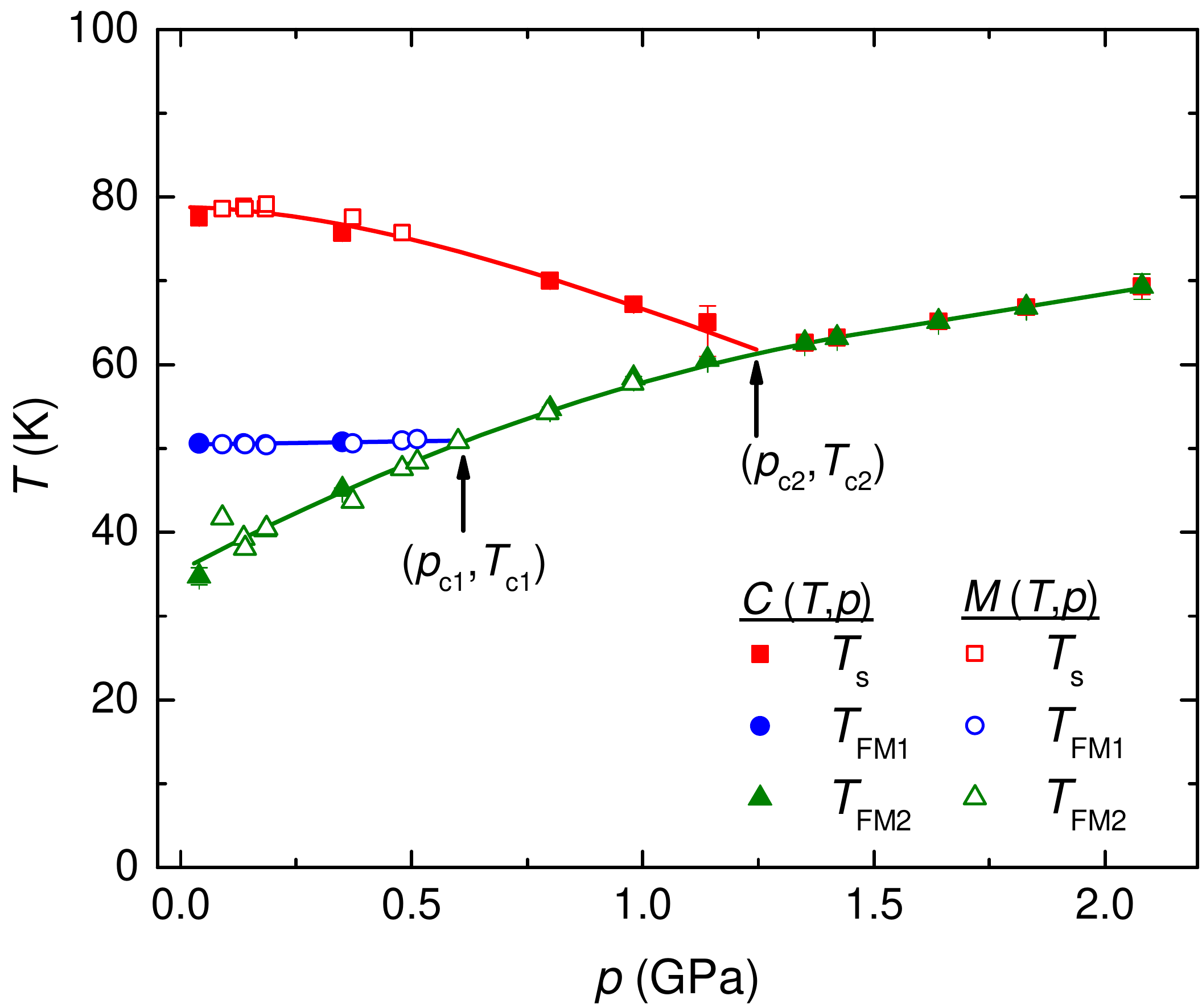} 
		\caption{Zero-field temperature ($T$)-pressure ($p$) phase diagram of VI$_3$, determined from specific heat measurements, $C(T,p)$ (full symbols), and magnetization measurements, $M(T,p)$ (open symbols). Squares denote the structural transition temperature $T_s$, circles the first ferromagnetic transition at $T_{FM1}$ and triangulars the second ferromagnetic transition at $T_{FM2}$. The arrows mark the position of the two critical points in the phase diagram at $(p_{c1},T_{c1})\,\approx\,(0.6\,\textnormal{GPa}, 50.8\,\textnormal{K})$ and $(p_{c2},T_{c2})\,\approx\,(1.25\,\textnormal{GPa}, 61.6\,\textnormal{K})$, which mark the merging of the two ferromagnetic transition lines and the ferromagnetic with the structural one, respectively. (Error for given pressure values is $\Delta p \,\approx\,\pm\,0.02$\,GPa, i.e. smaller than the data points.)}
		\label{fig:phasediagram}
		\end{center}
		\end{figure}

		Now we discuss the effect of pressure on each phase transition. The structural transition temperature $T_s$, which is located at $\approx\,78$\,K at ambient pressure, is monotonically suppressed upon increasing $p$ for $p\,\le\,1.14\,$GPa (see Figs.\,\ref{fig:CMlowp} (a-b,g) and \ref{fig:CMmedp} (a-b,f-g) for the $C/T$ and $M$ data sets and Fig.\,\ref{fig:phasediagram} (below) for the $T$-$p$ phase diagram). The corresponding specific heat feature becomes significantly smaller in size and more broadened. The peak shape speaks in favor of a discontinuous, first-order structural phase transition. This is consistent with conclusions from ambient-pressure studies which emphasize the sharp nature of the structural phase transition \cite{Tian19}. At $p\,=\,1.14\,$GPa, the specific heat feature at $T_s$ overlaps with the one of the FM2 transition (which will be discussed in all detail further below). As a result, the specific heat feature, associated with $T_s$ can only be identified as a high-$T$ shoulder at the FM2 peak (see Fig.\,\ref{fig:CMmedp}\,(g)), which does not give rise to a pronounced feature in d$(C/T)$/d$T$. For comparison, we included in Fig.\,\ref{fig:CMmedp}\,(g) an estimate of the specific heat peak, associated with the FM2 peak, at 1.14\,GPa as a dashed-dotted line. This estimate, which we denote by $(C/T)_{FM2}$, was obtained by shifting the FM2 specific heat peak at 0.98\,GPa, where $T_s$ and $T_{FM2}$ are well separated, in temperature and by subsequently renormalizing the specific heat value. The difference of the measured $C/T$ data and $(C/T)_{FM2}$ thus corresponds to an estimate of the specific heat, related to the structural transition at $T_s$, and is denoted by $(C/T)_s$ (see blue dashed-dotted line in Fig.\,\ref{fig:CMmedp}\,(g)). Indeed, the so-derived $(C/T)_s$ displays a broad maximum, similar to the data sets at lower pressures (0.8\,GPa and 0.98\,GPa). In analogy to the analysis of the latter data sets, we assign the maximum in $(C/T)_s$ at 1.14\,GPa to $T_s$. Nevertheless, we take the uncertainties of this procedure into account by assigning the $T_s$ value at 1.14\,GPa in the  $T$-$p$ phase diagram in Fig.\,\ref{fig:phasediagram} (below) a larger error bar. For even higher pressures ($p\,>\,1.14$\,GPa), the structural transition line likely merges with the FM2 transition line. We will discuss this aspect in all detail below, but first turn to a discussion of the FM1 and FM2 transition line at low pressures ($p\,\le\,1.14\,$GPa).

		For small pressures ($p\,\le\,0.51\,$GPa), $C/T$ and $M$ (see Fig.\,\ref{fig:CMlowp}\,(a) and (c)) each show indications for two separate ferromagnetic transitions at $T_{FM1}$ and $T_{FM2}$ (see Fig.\,\ref{fig:CMlowp}\,(e) for a plot of $C/T$ and $M$ data at very similar pressures ((0.35$\,\pm$0.01)\,GPa) on the same temperature scale). As shown in Figs.\,\ref{fig:CMlowp}\,(a-d), $T_{FM1}$ is almost unchanged with pressure. The specific heat anomaly at $T_{FM1}$ remains of $\lambda$-type, and therefore of second order. At the same time, $T_{FM2}$ is strongly increased with pressure and shifts towards $T_{FM1}$ rapidly (see phase diagram in Fig.\,\ref{fig:phasediagram} (below)). As a result, the increase of $M$ takes place in two steps which are almost indiscernible in $T$ at $p\,\approx\,$0.5\,GPa. In a previous work \cite{Son19}, this behavior was assigned to domain dynamics; our results of the specific heat clearly speak against this scenario: notably, the specific heat feature at $T_{FM2}$ evolves into a pronounced, sharp peak by applying modest pressures ($p$\,=\,0.35\,GPa) (see Figs.\,\ref{fig:CMlowp}\,(a) and (e)). This clearly strengthens the notion of two ferromagnetic phase transitions in VI$_3$ at low pressures. 
		
		The continuous increase in $M$ at $T_{FM2}$ at low pressures, together with the tiny, very broad feature in $C/T$, speak in favor of a second-order or very weak first-order phase transition. However, the situation becomes clearer upon increasing pressure  (but still in the regime $p\,\le\,0.51\,$GPa), as the features of the FM2 transition in $C/T$ and $M$ become significantly sharper. From this observation, it seems likely that the transition evolves into a first-order transition upon increasing pressure.

		For intermediate pressures, 0.6\,GPa$\,\le\,p\,\le\,$1.14\,GPa, we find only one anomaly in $C/T$ and $M$ (see Figs.\,\ref{fig:CMmedp}\,(a) and (c) and Fig.\,\ref{fig:phasediagram} (below) for the phase diagram), which can be associated with a ferromagnetic transition. At this transition, the magnetization increases steeply without any signatures of multiple transitions. The specific heat shows a slightly asymmetric, but sharp and pronounced peak at the same temperature (see Fig.\,\ref{fig:CMmedp}\,(e) for a plot of $C/T$ and $M$ on the same temperature scale at $\,\approx\,$0.98\,GPa), reminiscent of a near first-order phase transition. These data indicate that the system undergoes only a single magnetic transition in this pressure range which we label with $T_{FM2}$.
		
		At even higher pressures (for $p\,>\,1.14\,$GPa), the FM2 transition and the structural transition merge. These pressures above 1\,GPa exceed the maximum pressure capability of the magnetization setup, and we therefore restrict the discussion here to results of $C/T$, shown in Figs.\,\ref{fig:CMhighp} (a-d). For 1.35\,GPa$\,\le\,p\,\le\,$2.08\,GPa, $C/T$ shows a single anomaly, which is significantly reduced in size and broadened upon increasing pressure, over the investigated $T$ range. This feature is likely a result of a broadened first-order singularity. To investigate the magnetic character of this transition, we also studied the effect of magnetic field on the specific heat for highest pressure, i.e., $p\,=\,$2.08\,GPa (see Figs.\,\ref{fig:CMhighp}\,(e-f)). By increasing the field to $\mu_0 H\,=\,9\,$T, the specific heat anomaly shifts to higher $T$ and becomes significantly broadened, suggesting a sizable shift of entropy to higher temperatures. This behavior of $C/T$ is consistent with expectations for ferromagnetic ordering. 
		
		The inferred transition temperatures, $T_{FM1}$, $T_{FM2}$ and $T_s$, are summarized in the $T$-$p$ phase diagram in Fig.\,\ref{fig:phasediagram}. It does not only depict	the good agreement of transition temperatures, inferred from $M(T,p)$ and $C(T,p)$, but also highlights the existence of two triple points: at $(p_{c1},T_{c1})\,=\,$(0.6\,GPa, 50.8\,K), the two ferromagnetic transition lines $T_{FM1}(p)$ and $T_{FM2}(p)$ merge into a single ferromagnetic $T_{FM2}$ line, which in turn merges with the structural transition line $T_{s}(p)$ into a combined magneto-structural $T_{FM2}/T_s(p)$ line at $(p_{c2},T_{c2})\,=\,$(1.25\,GPa, 61.6\,K).
		
\begin{figure}
		\begin{center}
		\includegraphics[width=\columnwidth]{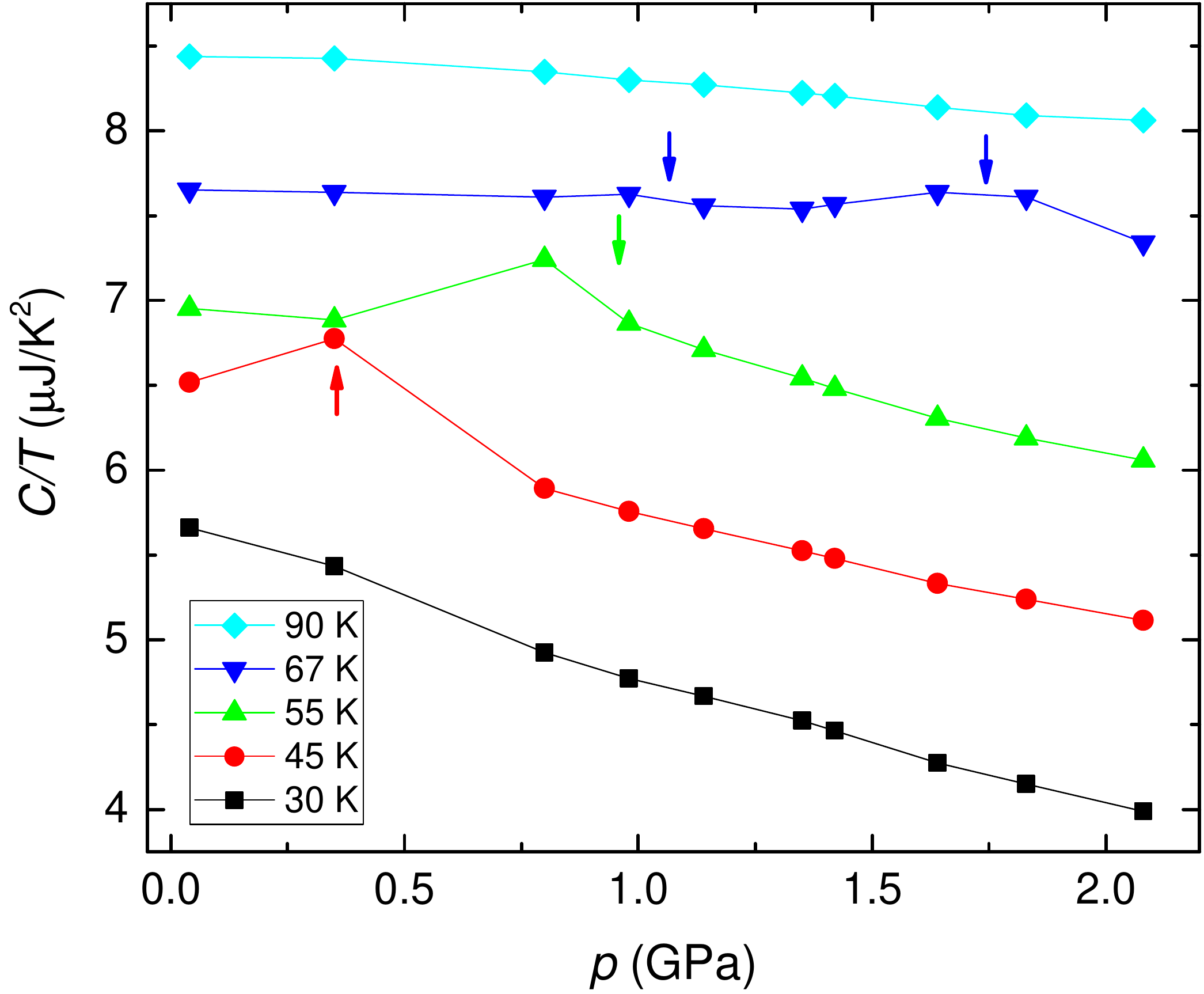} 
		\caption{Value of specific heat, $C/T$, vs. pressure, $p$ at constant temperatures $T\,=\,$30\,K, 45\,K, 55\,K, 67\,K and 90\,K. The arrows indicate the position of the various anomalies, associated with magnetic and structural transitions, in VI$_3$ under pressure (see main text for details). (Error for given pressure values is $\Delta p \,\approx\,\pm\,0.02$\,GPa, i.e. smaller than the data points.)}
		\label{fig:Cvsp}
		\end{center}
		\end{figure}

		In general, specific heat measurements alone do not allow for conclusions about the nature of a phase transition. Thus, it might also be possible that the single specific heat anomaly at $p\,\geq\,1.35$\,GPa results from a magnetic transition without any simultaneous structural transition. In this scenario, the crystallographic structure of VI$_3$ at low temperatures would be different for $p\,\le\,1.14\,$GPa and $p\,\geq\,1.35$\,GPa. This should, in principle, reflect itself in a feature in the pressure-dependent specific heat at fixed temperatures as a result of a change in the lattice specific heat. In Fig.\,\ref{fig:Cvsp}, we show a plot of $C/T$ values as a function of $p$ at five selected temperatures ($T\,=\,30\,$K, 45\,K, 55\,K, 67\,K and 90\,K). As can be seen by comparison to the phase diagram in Fig.\,\ref{fig:phasediagram}, the 90\,K cut should be featureless and, indeed, no feature is observed in the $C/T$ vs. $p$ plot within our resolution. In contrast, the $C/T$ data at $T\,=\,$67\,K, 55\,K and 45\,K display various features (marked by arrows), associated with crossing the appropriate $T_s$, $T_{FM1}$ and/or $T_{FM2}$ lines. The data at 30\,K again do not show any features within our experimental resolution. This observation suggests that VI$_3$ at low temperatures does not undergo any phase transition as a function of pressure. In turn, this implies that the phase transition at $p\,\geq\,1.35$\,GPa is indeed a combined magneto-structural transition. This picture only does not hold, if the change of entropy associated with this transition becomes so small that it falls below our resolution. For a definite proof of a merged magneto-structural transition at high pressures (for $p\,>\,1.4\,$GPa), x-ray and/or neutron scattering experiments at high pressures are desirable.
		
		\section{Discussion}

Prior to a discussion of the implications of our $T$-$p$ phase diagram, presented in Fig.\,\ref{fig:phasediagram}, we would like to point out that many of our results are qualitatively consistent with earlier reports. This relates to (i) multiple step-like features in $M(T)$ at ambient and at low pressures, as well as (ii) the transition temperatures. However, by combining the magnetization data with specific heat data under pressure and ambient-pressure NMR data, we reach different conclusions. First, the multiple step-like features in $M(T)$ correspond to separate ferromagnetic phase transitions. Second, as a consequence, the critical pressure $p_{c1}$ is not a result of a dimensional crossover. This scenario was suggested earlier \cite{Son19}, based on the break of the slope of the $T_{FM1}$ line and the $T_{FM2}$ line at $p_{c1}$. It was argued that the initial pressure insensitivity of $T_{FM1}$ speaks in favor of a two-dimensional ferromagnetism which becomes significantly three-dimensional only above $p_{c1}$. In contrast, our results identify $p_{c1}$ with a triple point at which two phase transition lines merge. 

Upon approaching the triple point at $p_{c1}$ from low pressures, we argued above that the high-$T$ transition at $T_{FM1}$ is a second-order transition, whereas the low-$T$ transition at $T_{FM2}$ is likely a first-order transition. Indeed, there are thermodynamic constraints on the order of the phase transition when they merge, as outlined by Yip \textit{et al.} \cite{Yip91}. Following these arguments, the low-$T$ transition must be a first-order transition, when approaching $p_{c1}$ from low pressures, irrespective of the order of the transition at $p\,>p_{c1}$. Our data does not allow for a conclusive statement of the character of the phase transition for $p_{c1}\,<\,p\,<\,p_{c2}$. In fact, our data is compatible with either a second-order transition which is almost first-order or with a weakly first-order transition. From a thorough analysis \cite{Liu19} of the critical exponents of the $T_{FM1}$ transition at ambient pressure from modified Arrott plots, it was concluded that this transition is best described by tricritical mean-field model, thus implying that it is locately closely to a point at which a change from second to first order occurs. This point can be accessed upon tuning by an external parameter, like field or pressure. Even though we cannot identify the triple point at ($p_{c1},T_{c1}$) as a tricritical point due to the lack of unambiguous assignment of the character of the the phase transition at $p\,>\,p_c$, our results still indicate that the character of the two transitions can be easily manipulated when applying modest pressures. This likely will reflect itself also in the critical exponents at ambient pressure, and is therefore at least consistent with the study of Liu \textit{et al.} \cite{Liu18}.

The present results call for an identification of the magnetic structure of VI$_3$ below $T_{FM1}$ and $T_{FM2}$, respectively. Despite our microscopic NMR study at ambient pressure, we cannot uniquely identify the ordering wavevectors below the two transitions. This is partially rooted in the fact that no consensus has been achieved yet on the crystal structure of VI$_3$ at room temperature and below the structural transition at $T_s$. Nonetheless, our NMR data at least allows for constraints on the types of magnetic order. In particular, this includes that the low-temperature magnetic order ($T\,<\,T_{FM2}$) has two sites with ordered V moments with an occupancy ratio of approximately 2:1. However, it is surprising that the hyperfine field on the V site is comparably small. Another peculiarity of the low-$T$ ordered state is that fields above 1\,T saturate the magnetization at $T\,=\,1.8\,$K \cite{Kong19,Son19,Liu19}, but that the magnetization value is significantly lower than the expectations for a $S\,=\,1$ system for V$^{3+}$. The latter two observations might indicate that orbital degrees of freedom might need to be taken into account in the description of the properties of VI$_3$. Another possibility might involve a canting of the moments in VI$_3$. Overall, to reconcile all these experimental findings, the magnetic structure remains to be verified by neutron scattering experiments and/or supported by theoretical calculations. In either case, as a first step prior to the determination of the magnetic structure, consensus has to be achieved on the crystallographic structure \cite{Kong19,Tian19,Son19} in all salient regions to identify V sites with different lattice symmetries. These future studies of the lattice properties (by e.g. x-ray scattering or thermal expansion) of VI$_3$ should also include very low temperatures, i.e., 30 K and below, as the first-order character of the $T_{FM2}$ transition at low pressures also suggests discontinuous lattice changes which potentially result in a different crystallographic symmetry.

Irrespective of the detailed magnetic structure, the herein determined $T$-$p$ phase diagram of VI$_3$ clearly demonstrates a strong coupling of the magnetic and structural degrees of freedom. A particular manifestation of this interplay is disclosed at the triple point $(p_{c2},T_{c2})$: the structural and magnetic transition, which occur for low pressures ($p\,<\,p_{c2}$) at very distinct ordering temperatures with $T_{FM2}\,<\,T_{s}$, can be tuned to a simultaneous first-order magneto-structural transition for $p\,>\,p_{c2}$ with $T_{FM2}\,=\,T_{s}$. A similar scenario, i.e., a triple point at which magnetic and structural transition lines merge, was also found in other compounds, such as MnNiGe \cite{Anzai78}, or even iron-based superconductors \cite{Fernandes14}. In fact, most of the van-der-Waals based magnets undergo a temperature-induced first-order structural transition \cite{McGurie17,Morosin64}, before developing long-range magnetic order upon cooling. Even though in most cases the response of the structural transition to pressure has not been studied to date, it seems plausible that this coupling of structural and magnetic degrees of freedom is a common feature in all these systems \cite{McGuire15}.

Another hallmark of the strong magneto-elastic coupling in this compound is the high sensitivity of the transition temperature $T_{FM2}$ to pressure. $T_{FM2}$ increases upon pressurization with d$T_{FM2}$/d$p\,=\,+(22\,\pm\,1)$\,K/GPa (this value corresponds to the average slope for $p\,<\,1\,$GPa). This pressure dependency is one order of magnitude larger than the one of $T_{FM1}$ (d$T_{FM2}$/d$p\,=\,+(1.1\,\pm\,0.3)$\,K/GPa) and distinctly larger than the ones found for other 2D van-der-Waals magnets, such as CrI$_3$ \cite{Mondal19}, CrBr$_3$ \cite{Yoshida97}, or Cr$_2$Ge$_2$Te$_6$ \cite{Sun18}.

 In Cr$_2$Ge$_2$Te$_6$, however, $T_c$ is actually decreased by application of pressure. There, the negative slope of $T_{FM}$ with $p$ was attributed \cite{Sun18} to combined pressure-induced changes of the direct Cr-Cr distance as well as the Cr-Te-Cr angle, corresponding to the superexchange interaction path. In particular, it was argued that the direct Cr-Cr distance decreases and that the Cr-Te-Cr angle is declined from 90$^\circ$. Both of these effects are argued to act in favor of antiferromagnetism and weaken ferromagnetism. If one applies this picture of Cr$_2$Ge$_2$Te$_6$ to VI$_3$, this might suggest that the positive pressure dependences in VI$_3$ can be attributed to pressure changes of the metal-ligand-metal angle. At the same time, one should keep in mind that pressure in these strongly anisotropic 2D van-der-Waals magnets will also significantly strengthen the inter-layer coupling, and as such, stabilize magnetism. To understand which of these factors actually play the decisive role at the first and second ferromagnetic transition in VI$_3$, studies of the structure under pressure are needed. Whatever changes in the structure under pressure in detail, it does affect the low-$T$ ferromagnetic transition at $T_{FM2}$ more than the high-$T$ ferromagnetic transition at $T_{FM1}$ for $p\,<\,p_1$, as displayed by a much stronger sensitivity to external pressure.

\section{Conclusions}

In summary, we presented a combined study of microscopic magnetic properties of the van-der-Waals magnet VI$_3$ at ambient pressure with thermodynamic properties (specific heat and magnetization) at finite pressures. These results show that at ambient pressure VI$_3$ undergoes a structural transition at $T_s\,\approx\,78\,$K and two distinct ferromagnetic transitions at ambient pressures with $T_{FM1}\,\approx\,50$\,K and $T_{FM2}\,\approx\,36\,$K. For $T\,\le\,T_{FM2}$, two ordered V sites exists with an occupancy ratio of approximately 2:1, whereas for $T_{FM2}\,\le\,T\,\le\,T_{FM1}$ only one ordered V site exist. Thus, VI$_3$ exhibits a complex magnetic structure. Under pressure, these two magnetic transitions merge at ($p_{c1},T_{c1}$)$\,\approx\,$(0.6\,GPa,\,50.8\,K), and this line merges at even higher pressures with the line of structural phase transitions at ($p_{c2},T_{c2}$)$\,\approx\,$(1.25\,GPa,\,61.6\,K). Therefore, VI$_3$ undergoes a simultaneous magneto-structural transition for $p\,>\,p_2$. This speaks in strong favor of magneto-elastic coupling being of significance for the magnetic properties in this compound. As a consequence, these results call for a clarification of the crystallographic and magnetic structures of this compound in all salient temperature regions.
%last, but not least: combination of spec heat and M useful tool to use the pressure phase diagrams of these semiconducting/insulating materials

\begin{acknowledgements}
Work at the Ames Laboratory was supported by the U.S. Department of Energy, Office of Science, Basic Energy Sciences, Materials Sciences and Engineering Division. The Ames Laboratory is operated for the U.S. Department of Energy by Iowa State University under Contract No. DEAC02-07CH11358. E.G. was partially supported by the Gordon and Betty Moore Foundation’s EPiQS Initiative through Grant GBMF4411. Work conducted at Princeton University was supported by the NSF-sponsored PARADIGM program centered at Cornell University, grant DMR-1539918. Y.I. thanks JSPS Program for Fostering Globally Talented Researchers which provided an opportunity to be a visiting scholar at Ames Laboratory.
\end{acknowledgements}	

\section{Appendix}

\subsection{Magnetic hysteresis measurements in the ferromagnetic state}	

\begin{figure}
		\begin{center}
		\includegraphics[width=0.8\columnwidth]{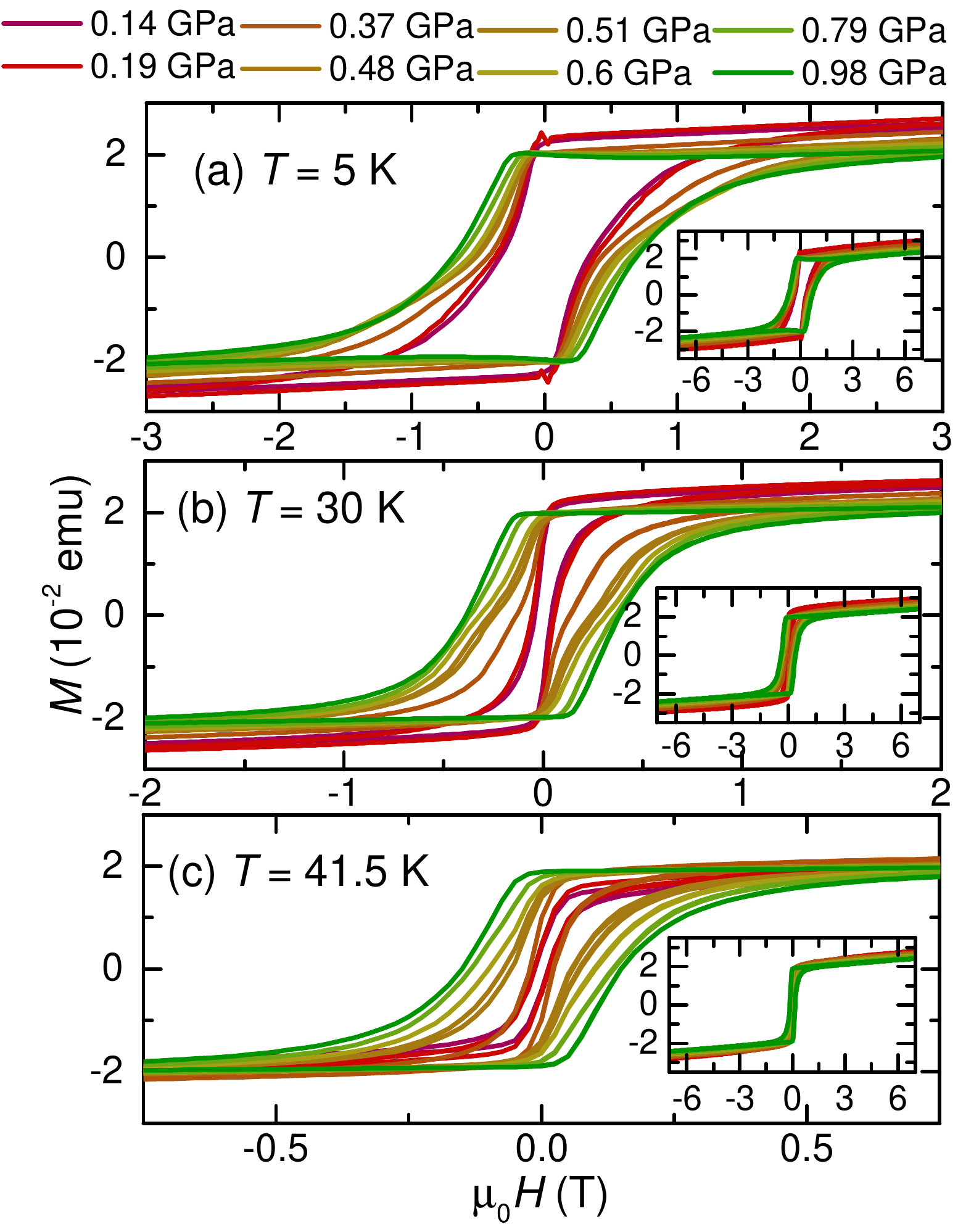} 
		\caption{Magnetization, $M$, vs. field, $\mu_0 H$, at $T\,=\,$5\,K (a), 30\,K (b) and 41.5\,K (c) at different pressures between 0.14\,GPa and 0.98\,GPa.}
		\label{fig:hysteresis}
		\end{center}
		\end{figure}

Field-dependent magnetization data under pressure were taken at three constant temperatures ($T\,=\,5\,$K, 30\,K and 41.5\,K) in fields up to $\pm\,7\,$T. At each temperature, the field was first increased from 0\,T to 7\,T, subsequently decreased to -7\,T and then increased to 7\,T. The results are shown in Fig.\,\ref{fig:hysteresis}. As these data were taken on a randomly-oriented aggregate of single crystals, the shown hysteresis loops are a superposition of loops parallel and perpendicular to the magnetic easy axis. At all temperatures and pressures, a pronounced hysteresis can be observed. This therefore shows that FM1 (represented by data taken at 5\,K and 30\,K and high-pressure data ($p\,>\,0.6\,$GPa) at 41.5\,K) as well as FM2 (represented by low-pressure data ($p\,<\,0.6\,$GPa) at 41.5\,K) are ferromagnetic in nature.

\bibliographystyle{apsrev}

\end{document}